\documentclass[a4paper,11pt]{article}
\usepackage{jinstpub} % for details on the use of the package, please see the JINST-author-manual
\usepackage{lineno}
\usepackage{natbib}[sort&compress]
\usepackage{fancyvrb}
\usepackage{tcolorbox}
\usepackage{multicol}
\usepackage[frozencache,cachedir=.]{minted}
\usepackage{xcolor} % to access the named colour LightGray
\definecolor{LightGray}{gray}{0.9}
\title{EPICS for Small-Scale Laboratories with Python Soft IOCs}

\author{J.~D.~Maxwell}
\affiliation{Thomas Jefferson National Accelerator Facility, Newport News, VA}

\emailAdd{jmaxwell@jlab.org}

\abstract{While the Experimental Physics and Industrial Control System (EPICS) is widely used at large laboratories for slow controls and instrumentation, the deployment of a full EPICS installation can be difficult, with a steep learning curve to new users. Taking advantage of the pythonSoftIOC module, we developed an EPICS slow controls implementation for Jefferson Lab's Hall B cryotarget written entirely in Python and based on software IOCs that communicate with instruments over Ethernet. This system ran successfully, interfacing with Jefferson Lab's full EPICS network, and we offer it as an example of the capabilities of pythonSoftIOC to build lightweight, yet robust and flexible instrumentation platforms that would be easily adapted for use at a small-scale laboratory. University groups can use these examples to build complete slow controls systems, from device communication to data archiving and display, using open-source, mature EPICS tools and student-friendly Python as an alternative to expensive and proprietary systems such as LabView.
}


\begin{document}
\maketitle
\flushbottom

\section{Introduction}
\label{sec:intro}
The Experimental Physics and Industrial Control System (EPICS) provides a crucial backbone of control and instrumentation for large-scale scientific efforts across the world, from particle accelerators to telescopes \cite{kozubal_run-time_1990, white_evolution_1999}. An EPICS system consists of many distributed computers using client-server and publish-subscribe techniques to coordinate data collection and control in real-time. Individual servers, called ``Input/Output Controllers'' or \textbf{IOCs}, interact with instruments to gather data---such as a temperature monitor reading out over a serial interface---or control actions---such as a relay controlling the current to a solenoid valve. IOCs can be hosted on standard PCs, but they are often on embedded system processors such as single-board computer in a VME crate. IOCs organize data in databases consisting of records called process variables (\textbf{PVs}) of many various types, such as analog input, binary output, state or waveform. Through these PVs, clients on the network can interact with devices on various IOCs~\cite{epics_epics_nodate}.
While EPICS is used at large laboratories world-wide and is a well-supported and featureful open-source project, these immense capabilities bring a degree of complexity that gives developing for EPICS a steep learning curve. EPICS is written in C/C++, and the application developer guide is over 300 pages long~\cite{noauthor_welcome_nodate}. 

\textbf{pythonSoftIOC} is a module developed at the Diamond Light Source in the UK which allows the creation of EPICS IOCs from directly within the Python interpreter~\cite{noauthor_pythonsoftioc_nodate}. This simplifies EPICS deployment and gives developers access to tools and instruments built to interact with one of the most popular programming languages. Python's clear syntax, dynamic typing and interpreter execution make it easy to learn as an introductory language, so laboratory students are often familiar with it. While these soft IOCs offer a limited subset of the features of a full EPICS system, the ability to dynamically create PVs and even full IOCs in a Python script adds valuable flexibility.

We were tasked with the implementation of EPICS control and data acquisition for a new cryotarget system in 2023. Jefferson Lab's Continuous Electron Beam Accelerator Facility (CEBAF) produces beams of electrons at up to 12\,GeV, which are used for basic nuclear and particle physics experiments in one of 4 experimental halls, A, B, C or D. The Hall B cryotarget used liquid helium from an end station refrigerator as a coolant to liquify gasses (typically H$_2$ or D$_2$), filling a kapton target cell in the interaction region of the detector system. The measurement of temperatures, pressures and flows in the system, as well as the control of heaters, valves and flow controllers, was crucial to the efficient operation of the target, as was the integration of these systems into CEBAF's overall control architecture. The safe delivery of beam to the hall depended on positive confirmation of the state of the target---empty or full---to machine control.

We chose the Hall B cryotarget as a testbed to try out a fully Python-based EPICS control scheme on a small scale. Rather than utilizing traditional, physical IOCs, we opted for a fully Ethernet-connected system where a collection of soft IOC applications would coordinate with each instrument over the network. A host PC located anywhere on the network would run these soft IOCs and serve as their connection to the larger EPICS network at the lab. 

In this paper, we will outline the Python-based EPICS system implemented in Hall B and offer it as an example to quickly and easily bring EPICS to smaller-scale projects in place of more expensive, proprietary tools. We have made this working example code available on GitHub as \texttt{softioc-toolkit}~\cite{maxwell_jdmaxsoftioc-toolkit_2025}.

\section{Soft IOCs}
Traditional IOCs run on a computer or embedded system which is physically connected to a number of instruments, managing data and control signals to and from those devices. The IOC application hosts a database of records, processes information and makes PVs available to other parts of the EPICS network. Soft IOCs are applications which perform the usual duties of an IOC, but exist on a server not locally connected to any instruments.

The increasingly common inclusion of Ethernet adapters directly into instruments and the availability of Ethernet DAC, ADC and DIO modules means that soft IOCs can serve a large range of devices over a network from a single server, without any physical connections. Legacy interfaces are easily adapted to Ethernet as well, using readily available serial-to-Ethernet and GPIB-to-Ethernet adapters.

Our pythonSoftIOC implementation uses only network-enabled instruments for monitoring and control through EPICS PVs. Each instrument in the experiment has its own soft IOC, creating and databasing only the data and control PVs for that one device on the network. Each of these soft IOCs is created by a Python script that generates them dynamically based on the contents of a configuration file. An additional soft IOC, called the IOC manager, was made to create control PVs to start and stop each device soft IOC, so that the full system is controlled over EPICS.

\subsection{Master IOC Template}

In our implementation, the creation of an instrument's soft IOC is handled by only a few python classes, making up as little as 100 lines of code. An IOC begins with the master IOC template script, \texttt{master\_ioc.py}, which uses the pythonSoftIOC module to set up functions common to all device soft IOCs and start the event loop to manage the asynchronous execution of tasks. 
 Concurrent event handling is implemented using \textbf{asyncio}, which runs processes asynchronously, preventing slow devices from blocking execution~\cite{noauthor_asyncio_nodate}. In general, \textit{reads} from devices are performed regularly based on a delay time in the configuration, and \textit{sets}, or writes to a device, are performed as a triggered event when a PV is changed in the EPICS system.

Upon execution, the Master IOC script creates an instance of the \texttt{DeviceIOC} class, passing along to the class the settings which will determine the behavior of the IOC and the instrument it will coordinate. The \texttt{DeviceIOC} class uses \textbf{importlib} to load an instance of an instrument-specific \texttt{Device} class, as set in the configuration file. Once the \texttt{Device} instance is made, the \texttt{DeviceIOC} instance creates a read-only heartbeat PV, and then sets the standard EPICS record fields for each PV based on the listings in the configuration file---such as ``DESC'' for a plain-text description and ``PREC'' for the precision, as detailed in the EPICS documentation~\cite{noauthor_welcome_nodate}.

\subsection{Device Interaction}

Each type of instrument has its own file in the ``devices'' directory, which typically contains two classes specific to that instrument: the \texttt{Device} and the \texttt{DeviceConnection}. The \texttt{Device} class coordinates the connection between the instrument and the EPICS system. This class creates all the PVs required by the IOC, but also tells the EPICS system how to read and control the instrument, connecting PVs to instrument commands. The primary methods used for this are \texttt{do\_reads()} and  \texttt{do\_sets()}, which contain commands to read and control the instrument, and set alarm codes if they fail.

The Python commands executed by the \texttt{Device} class are created in the \texttt{DeviceConnection} class. This class contains the actual connection to the device, often through \textbf{Telnet}, and commands to send to the instrument to perform any action, often in \textbf{SCPI}---the Standard Commands for Programmable Instruments. Taking a Lakeshore 218 temperature monitor as an example, its \texttt{DeviceConnection} class has method \texttt{read\_all()}, which writes the command ``\texttt{KRDG? 0}'' to the Telnet interface and reads back the response via a regular expression looking for 8 instances of ``\texttt{([+-]{\textbackslash}d+.{\textbackslash}d+)}'' for the 8 temperature channels. The \texttt{DeviceConnection} class is where instruments with bespoke Python communication modules would be implemented, such as a Zaber motor controller that we utilized with its \textbf{zaber\_motion} Python module~\cite{noauthor_welcome_nodate}. 

\begin{figure}
	\centering
		\includegraphics[width=\columnwidth]{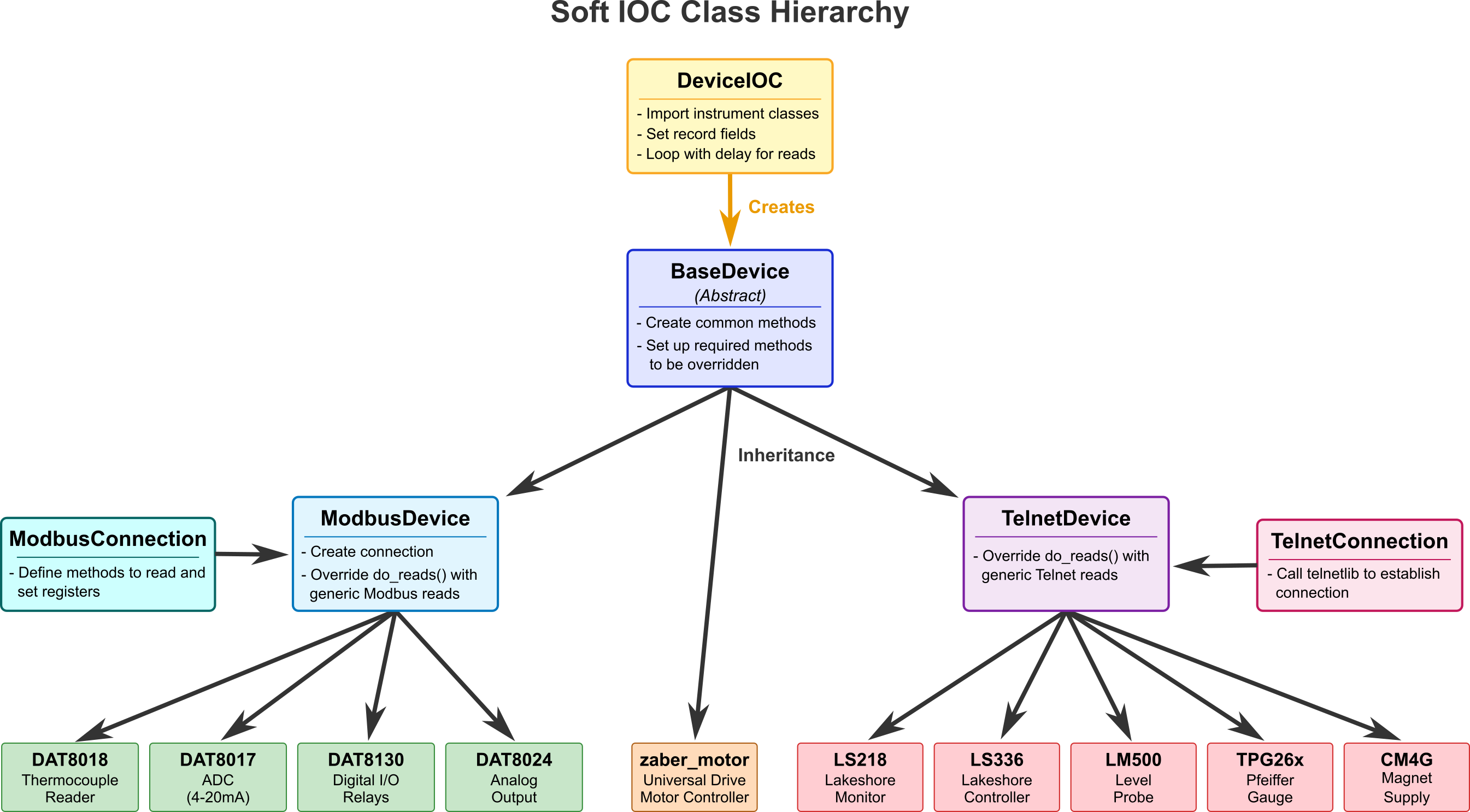}
	\caption{Device inheritance structure with example instruments.}
    
	\label{fig:inheritance}
\end{figure}

The simple class hierarchy of the \texttt{DeviceIOC} making the \texttt{Device} and the \texttt{Device} making the \texttt{DeviceConnection} is all that is needed for most instruments. In practice, much of the workings of the \texttt{Device} and \texttt{DeviceConnection} are common among many devices. To avoid repeated code and make it easier to maintain and extend, an abstract base class, \texttt{BaseDevice}, was made with default functions and blueprints for required methods which must be overridden in the child class. In addition, protocol specific parent classes---\texttt{TelnetDevice} and \texttt{ModbusDevice}---contain methods common to these protocols. Reading and writing Modbus registers and coils was particularly useful, as the Datexel company offers a number of analog input, analog output, digital in/out and relay devices that work on Modbus over Ethernet~\cite{noauthor_signal_nodate}. Figure \ref{fig:inheritance} shows the inheritance structure of the classes, with several example instruments inheriting from the abstract classes.

\section{Configuration}
%In traditional EPICS systems, a \texttt{.db} file directs the creation of PV records and fields for a given IOC, and these files can be complicated to a degree that there are stand-alone database configuration programs designed help users create them. 
In \texttt{softioc-toolkit}, a single, human-readable configuration file---formatted in YAML---directs the creation of both IOCs and PVs. With the exception of a general settings entry, each top-level key of this \texttt{settings.yaml} file defines a new IOC and gives all the necessary settings as a dictionary data structure to pass to each \texttt{Device} class. 

Figure \ref{fig:settings} shows an example configuration file entry for a Lakeshore 218 temperature monitor soft IOC. The device is given a name as the key for the entry, ``lakeshore\_218\_2,'' which in this case denotes the second instrument of this type in the system. The ``module'' key gives the location of the file containing \texttt{Device} and \texttt{DeviceConnection} for a Lakeshore 218, in this case a file called \texttt{ls218.py} in the ``devices'' directory. The network information of this instrument is given under ``ip'' and ``port;'' here the Lakeshore 218 is connected to a 4-port serial-to-ethernet adapter at ip address 192.168.0.111, and port 1002 is the second port on that device. The ``delay'' refers to how often the temperatures should be read from the monitor, in this case every 10 seconds. The ``records'' entry gives an optional listing of parameters, such as the plain-text description, precision, units or alarm limits, which set the standard EPICs fields of a PV.

\begin{figure}
	\centering
\begin{tcolorbox}[width=8cm]
    \begin{minted}{yaml}
lakeshore_218_2:
  module: 'devices.ls218'
  autostart: True      
  ip: '192.168.0.111'
  port: '1002'  
  timeout: 2
  delay: 10
  channels: 
    - Reservoir_TI
    - None
    - None
    - None
    - OVC_TI
    - Shield_TI
    - None
    - None
  records:
    Reservoir_TI:
      DESC: 'Reservoir Temperature'
      PREC: 2
      EGU: 'K'
      HIHI: 310
      HIGH: 300
      LOW: 2
      LOLO: 1
    Shield_TI:
      DESC: 'Shield Temperature'
      PREC: 2
      EGU: 'K'
      HIHI: 310
      HIGH: 300
      LOW: 2
      LOLO: 1
    \end{minted}
\end{tcolorbox}
	\caption{Example settings.yaml entry for a Lakeshore 218 temperature monitor.}   
	\label{fig:settings}
\end{figure}

The ``channels'' list refers to physical channels on the instrument; for a Lakeshore 218 there are 8 channels reading as many as 8 sensors. In this case, a single temperature input PV, denoted with a \texttt{\_TI}, is made for each channel. However, different \texttt{Device} classes could make a number of PVs for each channel, as with a magnet power supply where a ``Magnet'' listing would create \texttt{Magnet\_VI}, \texttt{Magnet\_Lead\_CI} and \texttt{Magnet\_ULIM} for the voltage, lead current and upper limit for that instrument. In the Lakeshore 218 example in Figure \ref{fig:settings}, three sensors are read into EPICS, on channels 1, 5 and 6, with ``None'' entries denoting when channels will not be used. The ``records'' listings flesh out the fields of the PVs, but they are not required, as seen with the missing entry for channel \texttt{OVC\_TI}.

\subsection{Logic Soft IOCs}
While most IOCs are tasked with coordinating an instrument for an experiment, they can also be used to monitor other IOCs' PVs, make calculations and perform actions. Python soft IOCs bring incredible flexibility for logic and analysis into an EPICS implementation, allowing data manipulation leveraging the full capabilities of Python's extensive library of modules and third-party add-ons. In the case of the Hall B cryotarget, it was necessary to set production states and report the target status to the accelerator machine control. For instance, if the operator wanted to fill the target with liquid hydrogen, a heater setpoint would be lowered to allow the target cell to cool, then raised slightly when the cell was sufficiently full. A logic soft IOC, \texttt{status\_ioc.py} was written to perform monitoring of pertinent PVs, change the states, and update control and alarm settings PVs for the target. 

Following the structure of the \texttt{Device} class, \texttt{status\_ioc.py} was executable and controllable in the same manner as all other instrument soft IOCs. The \texttt{do\_reads()} method was used to monitor PVs in the EPICS system to determine the production status and make changes to other PVs if required.
The key to reading EPICS PVs in time with other changes is the \textbf{aioca} module, an asynchronous input/output channel access client for asyncio, which allows EPICS gets and puts to be scheduled and run in asyncio.

A Proportional–Integral–Derivative (\textbf{PID}) controller is a commonly-used and often crucial feedback-based control loop. For the Hall B cryotarget, a heater in the main cryogen reservoir controlled the level of the liquid, and a PID loop determined the heater current necessary  to maintain a desired level. While pythonSoftIOC offers no PID record functionality, the direct access to Python within an IOC allows the implementation of PID modules or functions. We created a logic soft IOC to provide PID loops using the Python module \texttt{simple\_pid}, which operates in the same manner as other device IOCs. Any number PID loops, able to take any PVs on the network as input and output channels, can be created as stand-alone IOCs by making an entries to the \texttt{settings.yaml} file. % Might want to expand on this, comparing a python-based PID loop to the "standard" EPICS record for this.  Can be used to support the earlier statement(s) about learning curve.

\section{Managing IOCs}

Starting a soft IOC for an instrument is as simple as running a command line instruction: 
\begin{verbatim}
$python master_ioc.py -i lakeshore_218_2 
\end{verbatim}    
This starts a soft IOC for the second Lakeshore 218 in our lab. To instrument even a small lab, many IOCs may be needed, and starting, stopping or resetting these IOCs would become onerous over many terminal instances. To manage the operation of all the IOCs needed for the system, we created another soft IOC which spawns persistent processes to run the other soft IOCs.

Script \texttt{ioc\_manager.py} creates a single IOC which generates a \textbf{mbbOut} (multi-bit binary output) PV to control each soft IOC. In the case of our Lakeshore 218, the manager made a PV \texttt{lakeshore\_218\_2\_control} with three choices: \textit{stop}, \textit{run} or \textit{reset}. When the \textit{run} choice is set, the manager IOC spawns an instance of \textbf{screen}~\cite{screen_screen1_nodate} where it runs the command to start the instrument's soft IOC. In this way, all soft IOCs can be started and stopped from signals within the EPICS system. While \textbf{screen} has worked well to maintain many persistent terminals and their logs, we are investigating moving to Python's \textbf{supervisord} process control system, which offers simple controls and log access over the command line or a web server.

%Similarly to above, an opportunity to compare to other EPICS options.

\subsection{User Interface}
\begin{figure}
	\centering
		\includegraphics[width=\columnwidth]{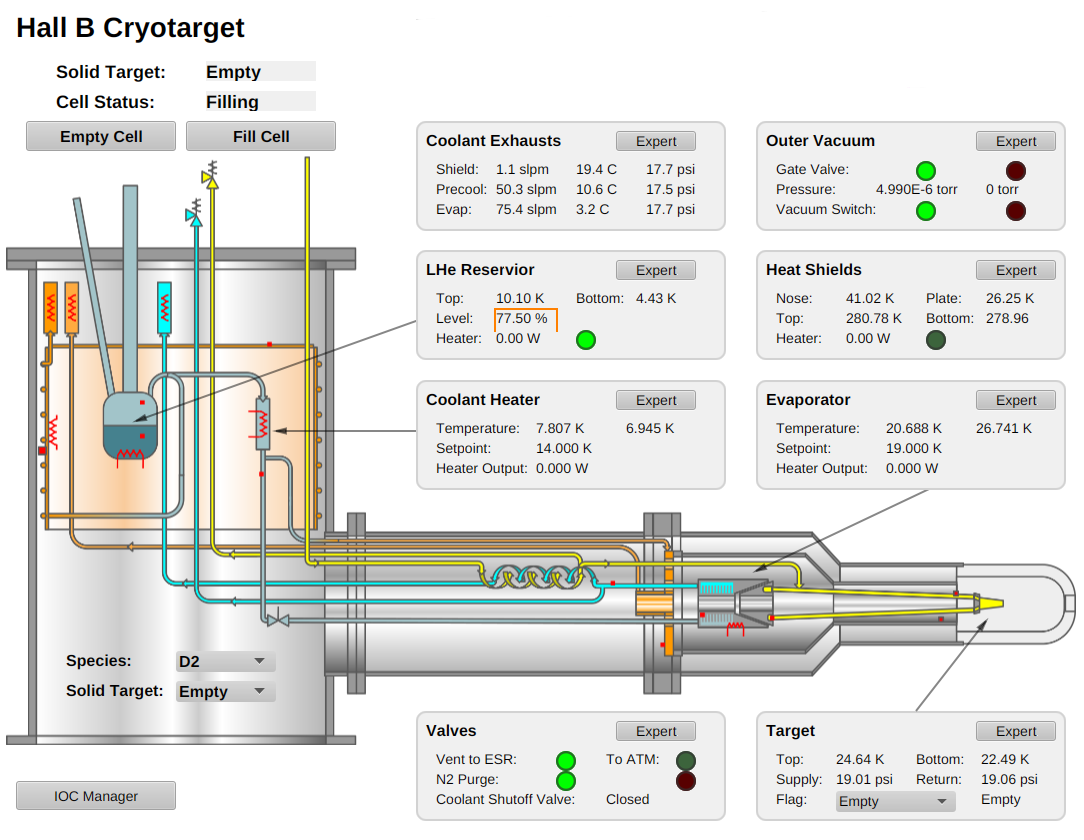}
	\caption{CS-Studio instrument panel interface for the Hall B Cryotarget, showing indicators and controls from numerous soft IOCs. }
    
	\label{fig:display}
\end{figure}

One benefit of using IOCs is the ease of producing a user interface with one of the mature front-end frameworks designed for use with EPICS. For the Hall B cryotarget, we used \textbf{CS-Studio/Phoebus}~\cite{noauthor_welcome_nodate}, which allows the drag-and-drop placement of control and display widgets to interact with EPICS PVs on the network. In addition to indicator elements showing the status of the system and controls to change the state, the soft IOC for every instrument in the system can be started and stopped from the GUI by the manager IOC. Because the manager starts a completely new process upon starting an IOC, any changes to the configuration file are loaded when restarting, so maintaining and updating soft IOCs is easy and accessible. Figure \ref{fig:display} shows the Hall B cryotarget panel in CS-Studio.

\subsection{Archiving and Display}
At large laboratories with full EPICS implementation, saving data is a vital task of the system  accomplished by sophisticated archiver appliances which can handle millions of PVs. The Hall B cryotarget took advantage of Jefferson Lab's archiver and data serving capabilities, the MySQL-based \textbf{Mya}~\cite{bickley_mysql-based_2008}. Access to such an archiving service might not be available at smaller labs, but would be a crucial element of a complete slow controls system. We developed a simple archiving utility, which saves PV values to CSV files over time, with appropriate deadband settings to record only when necessary. This archiver is run as another logic soft IOC, \texttt{archiver.py}, controlled like other devices by the usual ioc-manager PVs. For convenient access to the archived data, we also include an example of a simple web tool in \textbf{streamlit}~\cite{streamlit_streamlit_2021} to plot values from the archive over time.

\section{Conclusions}
Our implementation of Python soft IOCs for the new Hall B cryotarget was successful, and collaborators from Valparaiso University expanded upon it to instrument their solid target addition. 
Since the completion of the Hall B run with the new cryotarget in 2024, we have continued to utilize this Python-based EPICS framework in our development lab. We use a small, inexpensive, single-board computer running Ubuntu server to host all the IOCs, manager and archiver, and the front-end interface panel can be run on any machine on the network. The usability, expandability, and interoperability with the larger EPICS network at Jefferson Lab have encouraged us to share this toolkit with others. 

EPICS is an incredibly powerful and versatile framework, which supports a huge number of systems and instruments. Using Python soft IOCs as we have with \texttt{softioc-toolkit}, we are eschewing many of the advantages of existing support from the core EPICS software and its extensions, in favor of simplicity and flexibility while still maintaining compatibility with other EPICS systems. Although we've implemented only a few instruments thus far, because the interaction with each is transparent, adding new instruments should be accessible to novice coders, following our examples. The accessibility of the required Python modules with \textbf{pip} and our toolkit code through GitHub makes installation simple. The configuration and process management strategy we have followed adds to the flexibility of the system, making IOCs easy to create, run and maintain in experimental conditions.

While Python-based soft IOCs may not scale well to the millions of PVs needed at large labs, our \texttt{softioc-toolkit} offers easy-to-learn, easy-to-deploy slow controls examples for using EPICS at smaller labs or university groups. Groups that develop experimental sub-systems to bring to large labs will arrive with compatible software that easily interfaces with a full EPICS system. As free, open-source software, the toolkit could provide a less-expensive alternative to proprietary software and programming environments, written in a language that is widely-used and applicable across computer science.

\section*{Acknowledgements}
We gratefully acknowledge the advice and support of N. Baltzell. The Jefferson Lab target group scientists and technicians, especially J. Brock, deserve great credit for the design, construction and operation of the Hall B cryotarget on an incredibly tight schedule. This material is based on work supported by the U.S.~Department of Energy, Office of Science, Office of Nuclear Physics under contract DE-AC05-06OR23177.

\bibliographystyle{JHEP}
\bibliography{references}

\end{document}